\documentclass[prx,aps,amsfonts,floatfix,superscriptaddress,longbibliography,twocolumn,footinbib]{revtex4-2}
\usepackage[utf8]{inputenc}
\usepackage{amsmath}    
\usepackage{amssymb}
\usepackage{mathbbol}
\usepackage[dvipdfmx]{graphicx}   
\usepackage{bmpsize}
\usepackage{verbatim}   
\usepackage{subfigure}  
\usepackage{hyperref} 
\usepackage{scalerel}
\usepackage{cancel}
\usepackage{soul}
\usepackage{notes2bib}
\usepackage[normalem]{ulem}
\usepackage[dvipsnames]{xcolor}

\usepackage{siunitx}
\usepackage{physics}

\usepackage{enumerate}
\usepackage{empheq}
\definecolor{blue-violet}{rgb}{0.54, 0.17, 0.89}

\begin{document}

\title{Quantum tunneling and level crossings in the squeeze-driven Kerr oscillator}

\author{Miguel A. Prado Reynoso} 
\affiliation{Department of Physics, University of Connecticut, Storrs, Connecticut 06269, USA.}

\author{D. J. Nader}
\affiliation{Department of Chemistry, Brown University, Providence, Rhode Island 02912, United States}

\author{Jorge Ch\'avez-Carlos}
\affiliation{Department of Physics, University of Connecticut, Storrs, Connecticut 06269, USA.}

\author{B. E. Ordaz-Mendoza}
\affiliation{Department of Physics, University of Connecticut, Storrs, Connecticut 06269, USA.}

\author{Rodrigo G. Corti\~nas}
\affiliation{Department of Applied Physics and Physics, Yale University, New Haven, Connecticut 06520, USA}

\author{Victor S. Batista}
\affiliation{Department of Chemistry, Yale University, 
P.O. Box 208107, New Haven, Connecticut 06520-8107, USA}

\author{S.  Lerma-Hern\'andez}
\affiliation{Facultad  de F\'\i sica, Universidad Veracruzana,  Campus Arco Sur, Paseo 112, C.P. 91097  Xalapa, Mexico}

\author{Francisco P\'erez-Bernal}
\affiliation{Departamento de Ciencias Integradas y Centro de Estudios Avanzados en F\'isica, Matem\'aticas y Computaci\'on, Universidad de Huelva, Huelva 21071, Spain}
\altaffiliation[Also at ]{Instituto Carlos I de F\'{\i}sica Te\'orica y Computacional, Universidad de Granada, Fuentenueva s/n, 18071 Granada, Spain}

\author{Lea F. Santos}
\affiliation{Department of Physics, University of Connecticut, Storrs, Connecticut 06269, USA.}


\begin{abstract}
The quasi-energy spectrum recently measured in experiments with a squeeze-driven superconducting Kerr oscillator showed good agreement with the energy spectrum of its corresponding static effective Hamiltonian. The experiments also demonstrated that the dynamics of low-energy states can be explained with the same emergent static effective model. The spectrum exhibits real (avoided) level crossings for specific values of the Hamiltonian parameters, which can then be chosen to suppress (enhance) quantum tunneling. Here, we analyze the spectrum and the dynamics of the effective model up to high energies, which should soon be within experimental reach. We show that the parameters values for the crossings, which can be obtained from a semiclassical approach, can also be identified directly from the dynamics. Our analysis of quantum tunneling is done with the effective flux of the Husimi volume of the evolved states between different regions of the phase space. Both initial coherent states and quench dynamics are considered. We argue that the level crossings and their consequences to the dynamics are typical to any quantum system with one degree of freedom, whose density of states presents a local logarithmic divergence and a local step discontinuity. 
\end{abstract}

\maketitle


\section{Introduction}

Tunneling is a quantum mechanical phenomenon in which a system evolves through an energy barrier that it cannot penetrate classically. In the 1920s, tunneling successfully explained molecular spectral features~\cite{Hund1927a, Hund1927b}, electron emission from metals~\cite{Nordheim1928}, and $\alpha$-decay from nuclei~\cite{Gurney1929, Gamov1928}. Tunneling through inversion or internal rotation is deemed responsible for the details associated with high resolution molecular spectroscopy of equivalent conformers~\cite{KrotoBook}. A well-known example is the tunneling in the ground state of the ammonia molecule~\cite{KrotoBook,TownesBook}, whose associated energy splitting was used in the first maser,  pedagogically explained in~\cite{FeynmanBook}. Tunneling is also of major importance to understand chemical reaction rates, though the modeling of chemical reactions where tunneling plays a vital role is hampered by the high dimensionality of the systems involved and experimental difficulties~\cite{Meisner2016, Schreiner2020}. A recently published work has found a fine agreement between \textit{ab initio} calculations and experimental results for the tunnelling reaction of hydrogen molecules with deuterium anions~\cite{Wild2023}. There are several other observed examples of tunneling, such as in experiments involving atomic systems~\cite{Folling2007} and superconducting circuits~\cite{Devoret1985,Miller1994}. 

The literature is particularly vast for quantum tunneling in double-well systems \cite{Milburn1997, Leggett2001, Zollner2008, ZollnerPRA2008, Hunn2013, Halataei2017, Nader2021}, as in experiments with superconducting circuits~\cite{Harris2008}. Of particular interest to us is the case of driven oscillators, whose static effective Hamiltonians represent double-well systems and correspond to  period-doubling bifurcations~\cite{Dmitriev1986, Marthaler2006, Marthaler2007,dykman2012fluctuating}. Experimentally, this was recently achieved by applying a microwave drive to a superconducting nonlinear asymmetric inductive element (SNAIL) \cite{Frattini2017} transmon. 
The static effective Hamiltonian of this system corresponds to a squeeze-driven Kerr oscillator, where the squeezing amplitude and the detuning of the drive control the energy barrier of a double well~\cite{FrattiniPrep}. This implementation of a double-well system has the advantage of allowing in-situ tunability of all its Hamiltonian parameters and demonstrated access to the excited states~\cite{FrattiniPrep}, properties that we exploit in this work. Some of the dynamical properties of this system, which includes the exponential growth of the out-of-time ordered correlator due to instability~\cite{Pilatowsky2020}, were investigated in Ref.~\cite{ChavezARXIV}. Tunneling in the squeeze-driven Kerr oscillator was recently observed experimentally for the few lowest excited states in Ref.~\cite{Frattini2017,VenkatramanARXIV}. Here, we extend this study to higher energies.

The energy spectrum of the squeeze-driven Kerr oscillator presents real and avoided energy crossings for specific values of the Hamiltonian  parameters~\cite{VenkatramanARXIV}. They appear when the ratio $\Delta/K$ between the frequency $\Delta$ of the harmonic part of the Hamiltonian and the Kerr amplitude $K$ is an integer number~\cite{VenkatramanARXIV}. A similar behavior was theoretically predicted for the Lipkin-Meshkov-Glick  model~\cite{Nader2021}. Both systems have one degree of freedom, conserve parity, and are described by double wells. The crossings are real when the states have different parity, in which case tunneling is suppressed, and they are avoided for states with the same parity, leading to maximum tunneling. 

In the first part of this work, we provide a detailed analysis of the  Hamiltonian of the squeeze-driven Kerr  oscillator and employ semiclassical phase-space methods to elucidate the origin of the crossings. The values of the Hamiltonian parameters at which the crossings occur are determined via a semiclassical approach based on the Einstein-Brillouin-Keller quantization rule~\cite{Einstein1917, Brillouin1926, Keller1958}, as discussed in Refs.~\cite{Zhang2017, Nader2021, VenkatramanARXIV} (see also how to associate the crossings with the existence of a quasispin symmetry in~\cite{IachelloPrep}). We argue that these results are general to systems with one degree of freedom whose  density of states (DOS) exhibits a local logarithmic divergence and a local step discontinuity. 

In the second part of this paper, we use the effective flux of the Husimi volume to monitor the spread of an evolved state into specific regions of phase space that are classically forbidden. We explore two scenarios, one where the dynamics is initiated with a coherent state and the other where a Hamiltonian parameter is quenched. The tunneling amplitude changes depending on the ratio $\Delta/K$, decreasing (increasing) for values of $\Delta/K$ associated with real (avoided) crossings. We are able to show that the locations of the energy crossings can be dynamically identified.


\section{Model, Eigenvalues and Eigenstates}

The static effective Hamiltonian that describes the considered Kerr oscillator under a squeezing drive is written as~\cite{FrattiniPrep,VenkatramanARXIV}
\begin{equation}
\hat{H}= -\Delta \hat{a}^\dagger\hat{a} + K \hat{a}^{\dagger 2} \hat{a}^2 - \epsilon_2 (\hat{a}^{\dagger 2} +  \hat{a}^2), 
\label{Eqeff}
\end{equation}
where $\hat{a}^\dagger$ and $\hat{a}$ are the bosonic creation and annihilation operators, respectively, the amplitude $\Delta$ of the harmonic part of the Hamiltonian is the detuning between the frequency of the oscillator and half frequency of the drive, $K$ is the amplitude of the Kerr nonlinearity, $\epsilon_2$ is the squeezing amplitude,  and we set $\hbar =1$.  Hamiltonian (\ref{Eqeff}) conserves parity, that is $[\hat{ H}, {\cal \hat{P}}]=0$, where the parity operator ${\cal\hat{P}} = e^{-i\pi \hat{n}} = (-1)^{\hat{n}}$. We refer to the parity as either negative or positive. The notation for the eigenstates and eigenvalues is $\hat{H} |\psi_k \rangle = E_k |\psi_k \rangle$. 

The effective model in Eq.~(\ref{Eqeff}) is valid not only for small detuning values. Indeed, the experiments have so far reached $\Delta/K \sim 14$, and the model has remained valid. The experimental effective Hamiltonian in Ref.~\cite{VenkatramanARXIV} has an overall negative sign with respect to ours, $-\hat{H}$. We choose to write it as in Eq.~(\ref{Eqeff}) for convenience, so that we have wells instead of hills. In what follows, we take $\epsilon_2>0$. 

The Hamiltonian  in Eq.~(\ref{Eqeff}) is unbounded. We choose values for the size $N$ of the truncated Hilbert space  that guarantee the convergence of the eigenvalues and eigenstates under study. This unbounded Hamiltonian  exhibits several properties similar to those of the bounded Lipkin-Meshkov-Glick  Hamiltonian studied in \cite{Nader2021}. They include the presence of a quantum phase transition (QPT), one or two excited state quantum phase transitions (ESQPTs) \cite{Caprio2008,Stransky2016,Cejnar2021}, as well as real and avoided energy level crossings, which are the focus of this paper.

\subsection{Classical limit and stationary points}

\label{SecClass}

We gain insight over the system described by Eq.~(\ref{Eqeff}) by examining the classical limit \cite{ChavezARXIV, VenkatramanARXIV} of the Hamiltonian in the canonical variables $q$ and $p$, which reads
\begin{equation}
\begin{split}
H_{cl}
  &= -\frac{\Delta}{2}\left(q^2 + p^2\right) + \frac{K}{4} (q^2 + p^2)^2   -  \epsilon_2 (q^2  - p^2).
\end{split}
\label{EqHcl}
\end{equation}
The classical energy for a point $(q,p)$ in phase space is denoted by ${\cal E} = H_{cl} (q,p)$.

The first important step in our study of the classical Hamiltonian system is the identification of its stationary (critical) points. For the classical Hamiltonian (\ref{EqHcl}), the Hamilton’s equations of motion are 
\begin{equation}
\begin{split}
\dot{q} &= -p \left[
\Delta - K\left(q^2 + p^2 \right) - 2\epsilon_2\right], \\ 
\dot{p} &= \ q \left[
\Delta - K\left(q^2 + p^2 \right) + 2\epsilon_2\right].
\end{split}
\label{Hamil_eq}
\end{equation}
At the critical points we have that $\dot{q} = 0$ and $\dot{p}=0$.
Solving the system of equations above, we distinguish five critical points $r=\{q, p\}$, which are listed below together with their corresponding energy ${\cal E}$:
\begin{equation}
\begin{split}
& r_0=\{0,0\}, \hspace{2.6cm}   {\cal E}_{r_0}=0 ,\\
& r_1^{\pm}= \Big\{0, \pm\sqrt{\frac{\Delta-2\epsilon_2}{K}}\Big\}, \qquad 
{\cal E}_{r_1^{\pm}} = -\frac{\left(\Delta-2\epsilon_2\right)^2}{4K} ,\\ 
& r_2^{\pm}= \Big\{\pm\sqrt{\frac{\Delta+2\epsilon_2}{K}},0 \Big\},\qquad 
{\cal E}_{r_2^{\pm}} = -\frac{\left(\Delta+2\epsilon_2\right)^2}{4K}.
\end{split}
\end{equation}
By analyzing the stability of these points, we identify the following three scenarios~\cite{Nader2021,VenkatramanARXIV}. 

\begin{figure}[h!]
    \centering
    \includegraphics{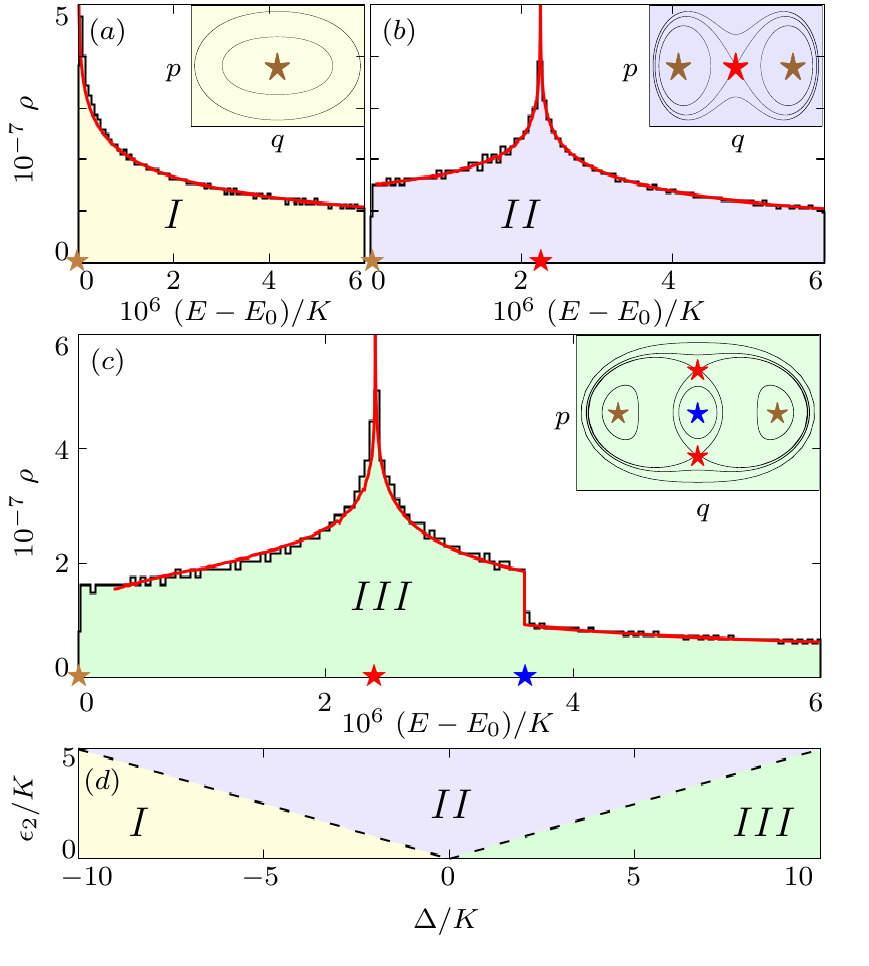}
\caption{Density of states [main panels (a)-(c)] calculated by exact numerical diagonalization, phase-space diagrams [insets (a)-(c)], and the parameter space (d) for regions I, II, and III. The red solid line in the main panels (a)-(c) indicates the semiclassical approximation of the DOS. The critical points of the phase-space diagrams in insets (a)-(c) and their energies in the main panels (a)-(c) are marked with stars: brown for minimum energy, red for unstable hyperbolic points, and blue for a local maximum. The DOS exhibits a peak (b) in Case (II) at $E^{\text{(II)}}_{{\rm{ESQPT}}}$ (red star), and both a peak at $E^{\text{(III)}}_{{\rm{ESQPT}}}$ (red star) and a step discontinuity at $E^{\text{(III)}}_{{\rm{step}}}$ (blue star) for the Case (III) displayed in (c). Case (I) in Fig.~\ref{fig01}~(a): $\Delta/K=-3000$ and $\epsilon_2/K=400$, Case (II) in Fig.~\ref{fig01}~(b): $\Delta/K=0$ and $\epsilon_2/K=1500$, and Case (III) in Fig.~\ref{fig01}~(c): $\Delta/K=3000$ and $\epsilon_2/K=400$.  The DOS includes both parities and is normalized considering the energy interval $E-E_0 \in [0,6\times10^6]$. The size of the truncated Hilbert space is $N=2000$ for each parity sector. }
\label{fig01}
\end{figure}

Case (I): For $\Delta/K \leq - 2\epsilon_2/K$, there is only one stationary point at $r_0$, as shown in the phase-space diagram of Fig.~\ref{fig01}~(a) (brown star). This stable point is a global minimum. 

Case (II): For $-2\epsilon_2/K< \Delta/K \leq 2\epsilon_2/K$, there are three stationary points, as shown in the phase-space diagram of Fig.~\ref{fig01}~(b). Two of them are stable points at $r_2^{\pm}$, which are global minima (brown stars), and $r_0$ now becomes an unstable hyperbolic point (red star).

Case (III): For $\Delta/K >2\epsilon_2/K$, there are five stationary points, as shown  in the phase-space diagram of Fig.~\ref{fig01}~(c). These are the same two stable global minima at $r_2^{\pm}$ (brown stars) as in Case (II), there are also two unstable hyperbolic points at $r_1^{\pm}$ (red stars), and $r_0$ recovers its status of a stable point, as in Case (I), though now it is a local maximum (blue star).

Figure~\ref{fig01}~(d) depicts the parameter space of the classical Hamiltonian in Eq.~(\ref{EqHcl}), classified in accordance with the three types of phase-space diagrams discussed above. 

\subsection{Density of states}

The DOS obtained with the eigenvalues $E_k$ of the Hamiltonian in Eq.~(\ref{Eqeff}) detects excited state quantum phase transitions associated with bifurcations \cite{dykman2012fluctuating} in the classical phase space~\cite{Stransky2016,Cejnar2021}, as we describe in this subsection. The main panels in Figs.~\ref{fig01}~(a)-(c) present the DOS as a function of the excitation energy $E' = E- E_{0}$, where $E_{0}$ is the ground-state energy. The shape of the DOS can also be reproduced from the semiclassical approximation, 
\begin{equation}
\rho\left({\cal E} \right)=\frac{1}{2\pi} \int dq dp \ \delta\left( H_{cl}-{\cal E} \right),
\label{Gut_trace}
\end{equation}
where $H_{cl}$ is the classical Hamiltonian in Eq.~(\ref{EqHcl}), and the result for Eq.~(\ref{Gut_trace}) is obtained using the lowest-order term of the Gutzwiller trace formula~\cite{GutzwillerBook}. This result is shown with a red line in the main panels of Figs.~\ref{fig01}~(a)-(c). Each one of the three cases identified in Sec.~\ref{SecClass} is characterized by a different structure of the DOS, as follows.

Case (I) in Fig.~\ref{fig01}~(a): The DOS decays monotonically as energy increases. A QPT happens at $\Delta/K = -2\epsilon_2/K$. This transition was studied in~\cite{Grigoriou2023} and can be understood from the analysis of the stationary points. As $\Delta/K$ grows from $-\infty$ and reaches $\Delta/K = -2\epsilon_2/K$, the single stationary point at $r_0$ duplicates into $r_2^{\pm}$. This implies that,  in the quantum domain, the energy difference between the ground state and the first excited state vanishes exponentially due to tunneling. 

Case (II) in Fig.~\ref{fig01}~(b): The DOS exhibits a peak at the energy denoted by $E^{\text{(II)}}_{\text{ESQPT}}$ (red star in the main panel), 
which  converges to the energy of the unstable critical point $r_0$ in the classical limit, 
\begin{equation}
E'^{\text{(II)}}_{\text{ESQPT}} \rightarrow {\cal E}_{r_0} - {\cal E}_{r_2^{\pm}} = \frac{\left(\Delta+2\epsilon_2\right)^2}{4K}.
\end{equation}
This accumulation of eigenvalues around $E'^{\text{(II)}}_{\text{ESQPT}}$ diverges as we approach the classical limit and is the main signature of what became known as ESQPT~\cite{Caprio2008,Cejnar2021}. The ESQPT critical energy value marks the boundary between two regions with different dominating symmetries. Pairs of eigenvalues, one with negative parity and the other with positive parity, are degenerate for energies below the ESQPT  energy, and they split once they get above the ESQPT energy~\cite{ChavezARXIV}.

Case (III) in Fig.~\ref{fig01}~(c): The DOS exhibits two distinct non-analytical features, a logarithmic peak and a discontinuous step. The peak is associated with the hyperbolic points (red stars in the inset) in the classical phase space and characterizes a first ESQPT. It happens at the energy $E^{\text{(III)}}_{\text{ESQPT}}$ (red star in the main panel), 
which converges to the energy of the unstable critical points  $r_2^{\pm}$ in the classical limit. Therefore, the excitation energy is given by
\begin{equation}
E'^{\text{(III)}}_{\text{ESQPT}} \rightarrow  {\cal E}_{r_1^{\pm}} - {\cal E}_{r_2^{\pm}}  =  \frac{2\Delta\epsilon_2}{K}.
\label{Eq:ESQPTiii}
\end{equation} 
The discontinuous step happens at an energy that converges to the energy of the local maximum (blue star in the inset) at the origin of the classical phase space,
\begin{equation}
E'^{\text{(III)}}_{\text{step}} \rightarrow  {\cal E}_{r_0} - {\cal E}_{r_2^{\pm}} = \frac{\left(\Delta+2\epsilon_2\right)^2}{4K}.
\label{Eq:stepiii}
\end{equation} 
The discontinuity at the energy $E^{\text{(III)}}_{\text{step}}$ (blue star in the main panel) characterizes a second kind of ESQPT. 

The dynamical consequences of the presence of the ESQPT in Case (II) for $\Delta=0$ were studied in \cite{ChavezARXIV}. In the present work, we focus in Case (III), where $\Delta/K > 2\epsilon_2/K$ and two ESQPTs exist; one associated with the two hyperbolic points, and the other with the local maximum.

\subsection{Avoided and real level crossings}

We now fix a value of $\epsilon_2/K$ and calculate the spectrum of the Hamiltonian (\ref{Eqeff}) as a function of $\Delta/K$ for Case (III). But before presenting these results, it is informative to analyze the trajectories of the underlying classical system (\ref{EqHcl}). 

\begin{figure}[h!]
    \centering
    \includegraphics{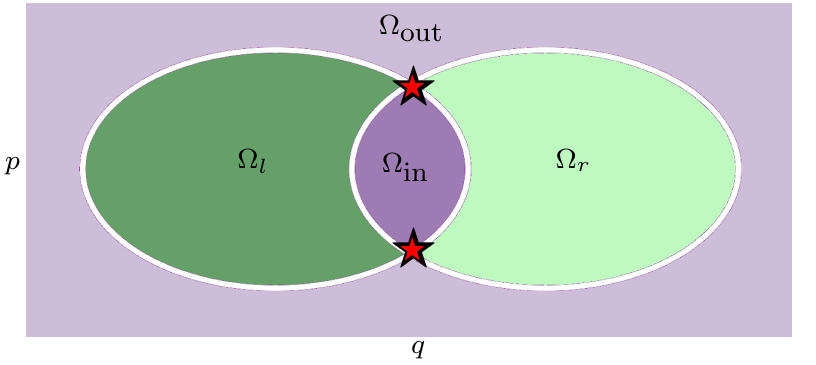}
\caption{Sketch of the classical phase space for Case (III) separated in regions defined by the separatrix (white color) of the first ESQPT, which is associated with the hyperbolic points (red stars). The regions $\Omega_{\text{l}}$ and $\Omega_{\text{r}}$ (green colors) contain pairs of degenerate classical trajectories with energies below the energy of the first ESQPT, ${\cal E}< E^{\text{(III)}}_{\text{ESQPT}}$. The regions $\Omega_{\text{in}}$ and $\Omega_{\text{out}}$ (purple colors) contain pairs of degenerate classical trajectories with energies between the first and second ESQPT, $E^{\text{(III)}}_{\text{ESQPT}}  < {\cal E}< E^{\text{(III)}}_{\text{step}}$.}
\label{fig02}
\end{figure}

Recall that the classical phase space of Case (III) has a local maximum at $r_0=\{0,0\}$ with energy ${\cal E}_{r_0}=0$. The trajectories with energy less than ${\cal E}_{r_0}$ are degenerate in pairs, that is, there are two trajectories with the same energy. These pairs of trajectories can be distinguished depending on the region of the phase space where they occur. One type belongs to the left and right regions of the phase space denoted as $\Omega_{\text{l}}$ and $\Omega_{\text{r}}$ in Fig.~\ref{fig02}. The other type is found in the $\Omega_{\text{in}}-\Omega_{\text{out}}$ regions of Fig.~\ref{fig02}. These two types of trajectory pairs manifest themselves differently in the quantum model, as described below.
 
(1) The $\Omega_{\text{l}}-\Omega_{\text{r}}$ region depicted in green in Fig.~\ref{fig02} contains pairs of degenerate trajectories with energy ${\cal E}$ less than the energy of the hyperbolic point,  $${\cal E}< {\cal E}_{r_1^{\pm}} \simeq E^{\text{(III)}}_{\text{ESQPT}}.$$  Each trajectory of the degenerate pair is located in one of the two wells centered at the global minima at $r_2^{\pm}$, with each one being the reflection in the $q$-axis of the other one. 

(2) The $\Omega_{\text{in}}-\Omega_{\text{out}}$ region depicted in purple in Fig.~\ref{fig02} contains pairs  of degenerate trajectories that have energy ${\cal E}$ between the energy of the hyperbolic points and the local maximum, $$E^{\text{(III)}}_{\text{ESQPT}} \simeq {\cal E}_{r_1^{\pm}} < {\cal E}< {\cal E}_{r_0} \simeq E^{\text{(III)}}_{\text{step}}.$$ For this region, contrary to what happens in the $\Omega_{\text{l}}-\Omega_{\text{r}}$ region, the size of the phase-space area covered by each trajectory of the degenerate pair is different, the one in $\Omega_{\text{in}}$ being smaller  than the one in $\Omega_{\text{out}}$.

\begin{figure}[h!]
    \centering
    \resizebox{1.\linewidth}{!}{\includegraphics{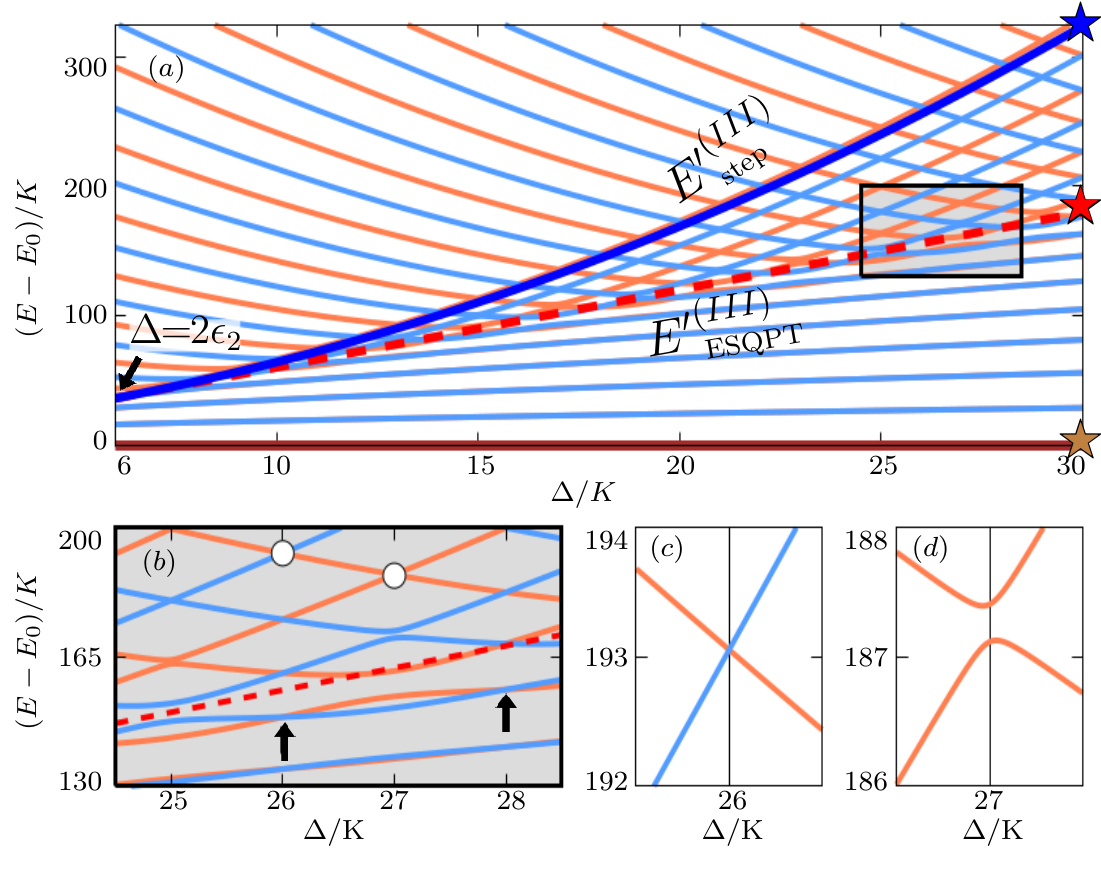}}
\caption{Excitation energy spectrum as a function of the control parameter $\Delta/K$ for fixed $\epsilon_2/K=3$ and $\Delta/K \geq 2\epsilon_{2}/K$. The plot corresponds to Case (III), where two ESQPTs exist [cf. Fig.~\ref{fig01}~(c)]. Orange (blue) lines indicate the levels in the positive (negative) parity sector. The red dashed line in panels (a)-(b) marks the energy ${E'}^{\text{(III)}}_{\text{ESQPT}}$ of the first ESQPT 
and the dark blue line in panel (a) corresponds to the energy  ${E'}^{\text{(III)}}_{\text{step}}$ of the second ESQPT.   
The small rectangle in panel (a) is amplified in panel (b). In the latter, the vertical arrows at $\Delta/K=26$ and $\Delta/K=28$ mark points where the degeneracy is not lifted as one approaches $E'^{\text{(III)}}_{\text{ESQPT}}$ from below, the circle at $\Delta/K=26$ indicates a real crossing that is shown again in panel (c), and the circle at $\Delta/K=27$ corresponds to an avoided crossing, which is evinced in panel (d).
}
\label{fig03}
\end{figure}

We are now in a position to further analyze the spectrum of the Hamiltonian (\ref{Eqeff}) for Case (III).
In Fig.~\ref{fig03}, we fix $\epsilon_2/K=3$ and show the eigenvalues as a function of $\Delta/K$. The results belong to Case (III), because the levels are shown for $\Delta/K>6$. The orange (blue) lines correspond to the energies in the positive (negative) parity sector. The red dashed line marks the first ESQPT energy, $E'^{\text{(III)}}_{\text{ESQPT}}$ [see Eq.~(\ref{Eq:ESQPTiii})], and the dark blue line shows the energy of the second ESQPT, associated with the discontinuous step, $E'^{\text{(III)}}_{\text{step}}$ [see Eq.~(\ref{Eq:stepiii})]. 

Considering Fig.~\ref{fig03}~(a), we can better understand the  features of the DOS observed in Fig.~\ref{fig01}~(c) and extract further details, as described next. 

$\bullet$ For ${E'}<{E'}^{\text{(III)}}_{\text{ESQPT}}$ [below the red dashed line in Fig.~\ref{fig03}~(a)], the spectrum is exponentially quasi-degenerate in pairs (kissing) for any value of $\Delta/K$. One level of the pair belongs to the positive parity sector and the other one to the negative parity sector, so the blue and orange lines are superposed. Classically, this is manifested as pairs of phase space trajectories with equal energy, where one trajectory belongs to the $\Omega_{\rm{l}}$ region and the other to the $\Omega_{\rm{r}}$ region of Fig.~\ref{fig02}. 

$\bullet$ As one approaches ${E'} \sim {E'}^{\text{(III)}}_{\text{ESQPT}}$ from below, the degeneracy between states of negative and positive parity is lifted throughout, except where $\Delta/K$ is even, as indicated with vertical arrows in Fig.~\ref{fig03}~(b).  
This is an important region of the spectrum for the analysis of quantum tunneling, as it will be explained in Sec.~\ref{Sec:EVEN}.

$\bullet$ Another especial spectral region for this work takes place at energies between the two ESQPTs of Case (III), 
\begin{equation}
E^{\text{(III)}}_{\text{ESQPT}}<E <E^{\text{(III)}}_{\text{step}} ,   
\label{Eq:box}
\end{equation}
that is, between the red dashed and dark blue lines in Fig.~\ref{fig03}~(a). Classically, this energy interval is associated with trajectories in the regions $\Omega_{\rm{in}}$ and $\Omega_{\rm{out}}$ of Fig.~\ref{fig02}. As seen in Figs.~\ref{fig03}~(a)-(b), there are crossings of energy levels for certain values of the control parameter $\Delta/K$.  They happen between energies of different parity sectors when $\Delta/K$ is even, and between energies within the same parity when $\Delta/K$ is odd. From a closer look at Figs.~\ref{fig03}~(c)-(d) and in agreement with the von Neumann-Wigner theorem~\cite{vonNeuman1929,Denkov2007}, one sees that the crossings for levels of different parities are real, while the crossings for levels of the same parity sector are avoided.

$\bullet$ For $E' > {E'}^{\text{(III)}}_{\text{step}}$ [above the dark blue line in Fig.~\ref{fig03}~(a)], the spectrum ceases to present any crossing, the energies simply increase alternating between positive and negative parity.

Motivated by the analysis in Ref.~\cite{Nader2021} (see also the Supplemental Material in Ref.~\cite{VenkatramanARXIV}), we show that the crossings in the energy region between the two ESQPTs, $E^{\text{(III)}}_{\text{ESQPT}}<E<E^{\text{(III)}}_{\text{step}}$, can be explained using a semiclassical approach based on the Einstein-Brillouin-Keller quantization rule, 
\begin{equation}
I = \frac{1}{2\pi} \oint p dq = \left(n + \frac{\mu}{4} + \frac{b}{2} \right),
\label{EBKrule}
\end{equation}
which applies to non-chaotic systems \cite{Keller1958}. In the equation above, $I$ is the action-angle coordinate for our system with one degree of freedom, and $n$ is a positive integer. The Maslov indexes, $\mu$ and $b$, indicate, respectively, the number of classical turning points in the trajectory  and the number of reflections with a hard wall. The trajectories for Hamiltonian~(\ref{EqHcl}) in the regions $\Omega_{\rm{in}}$ and $\Omega_{\rm{out}}$ have two turning points, so $\mu=2$, and no reflection with a hard wall, so $b=0$.  

To solve Eq.~(\ref{EBKrule}), we consider a canonical transformation to variables $(z,\phi)$ defined as $q=\sqrt{2z}\sin \phi$ and $p=\sqrt{2z}\cos \phi$, with $\phi$ covering the entire interval $\phi=[0,2\pi)$, and use the classical Hamiltonian in Eq.~(\ref{EqHcl}) to write
\begin{equation}
{\cal E} = H_{cl}(z,\phi) = K z^2 
+\left(\gamma-\Delta\right) z ,
\end{equation}
where $\gamma = 2\epsilon_2 \cos(2\phi)$.
For ${\cal E}\in [{\cal E}_{{r_2^{\pm}}},{\cal E}_{r_0}]$, the two solutions are
\begin{equation}
z_\pm({\cal E},\phi)=
\frac{\Delta - \gamma \pm \sqrt{(\Delta-\gamma)^2 +4K{\cal E}}}{2K}.
\end{equation}
Resorting to the Einstein-Brillouin-Keller quantization rule in Eq.~(\ref{EBKrule}) in terms of the canonical variables ($z,\phi$),
we obtain the semiclassical approximation for the energy of each trajectory,
\begin{equation}
\frac{1}{2\pi}\int_0^{2\pi}   
z_\pm({\cal E}_{n_{\pm}},\phi) d\phi = \left( n_\pm +\frac{1}{2}\right),
\label{EBKrulezphi}
\end{equation}
where ${\cal E}_{n_\pm}$ are the semiclassical energies and $n_\pm$ are integer numbers. If the energy levels are degenerate, ${\cal E}_{n_+}={\cal E}_{n_{-}}$, then Eq.~(\ref{EBKrulezphi}) gives
$$
\frac{1}{2\pi}\int_0^{2\pi}  (z_+ + z_{-}) d\phi = \frac{\Delta}{K}= \left( n_+ + n_{-} +1\right),
$$
which implies that
\begin{equation}
\label{AvoidingCrosingCondigion}
\frac{\Delta}{K}= n_+ + n_{-} +1.
\end{equation}
From this semiclassical approach, we conclude that there are crossings whenever $\Delta/K$ is an integer, although it is not possible to say whether the crossings are real or avoided. The distinction between the two is made in this work with the numerical diagonalization, as shown in Fig.~\ref{fig03} (see also \cite{VenkatramanARXIV} and \cite{IachelloPrep}), and they follow the von Neumann-Wigner theorem.


In a nutshell, the semiclassical approximation reveals that the degeneracies below the first ESQPT (below $E^{\text{(III)}}_{\text{ESQPT}}$) and the crossings between the two ESQPTs (between $E^{\text{(III)}}_{\text{ESQPT}}$ and $E^{\text{(III)}}_{\text{step}}$) are a consequence of the presence of at least two orbits in phase space with the same energy. If these classical orbits enclose identical areas (as in the $\Omega_{\text{l}}$ and $\Omega_{\text{r}}$ regions), then the corresponding quantum energies are degenerate. If the areas enclosed by two orbits with the same energy are different (as in the $\Omega_{\text{in}}$ and $\Omega_{\text{out}}$ regions), then the spectrum exhibits crossings for specific values of $\Delta/K$. The crossings are real for levels in different parity sectors and avoided for levels with the same parity.

We conjecture that these crossings at an intermediate energy region of the spectrum  emerge for any system with one degree of freedom, whose DOS is similar to the one for Case (III) seen in Fig.~\ref{fig01}~(c), that is, the DOS exhibits a local logarithmic divergence and a local step discontinuity, which must be consecutive and can appear in any order. The crossings are expected to appear at energies located between these two ESQPTs. Correspondingly, in the underlying semiclassical system, there should be at least two local stationary points in the phase space, a local hyperbolic point and a local center point, which are, respectively, related to the logarithmic divergence and the step discontinuity of the DOS. Other one-body systems that exhibit a similar spectral structures and energy crossings include the one-dimensional asymmetric double well case~\cite{Khuat1993} and, in many-body models with all-to-all couplings that can be effectively reduced to one degree of freedom, we have the Lipkin-Meshkov-Glick model~\cite{Nader2021} and the Bose-Hubbard model~\cite{Graefe2007}.


\subsection{Structure of the eigenstates}

The structure of the eigestates reflects the presence of the real and avoided crossings in the energy interval $E^{\text{(III)}}_{\text{ESQPT}}<E<E^{\text{(III)}}_{\text{step}}$ of the spectrum of Case (III). This
can be visualized by inspecting the Husimi functions of the eigenstates.
For an eigenstate $\ket{\psi_k}$, the Husimi function,  
\begin{equation}
  \mathcal{Q}_{\psi_k}^{(\alpha)} = \frac{1}{\pi}\left| \bra{\alpha}\ket{\psi_k}\right|^2
  = \frac{1}{\pi}\left| \sum_{n=0}^{N-1} c_{n}^k e^{-|\alpha|^2/2}\frac{{\alpha^*}^n}{\sqrt{n!}}\right|^2,
\label{Eq:HusimiStatic}
\end{equation} 
is defined as the absolute square of the projection of the eigenstate in the Glauber coherent state,
\begin{equation}
|\alpha\rangle = e^{-\frac{1}{2} |\alpha|^2} \sum_{n=0}^{N-1} \frac{\alpha^n}{\sqrt{n!}} |n \rangle ,
\label{EqSM_Coh}
\end{equation}
where  $\hat{a} |\alpha\rangle = \alpha|\alpha\rangle$, 
\begin{equation}
    \alpha = \sqrt{\frac{ 1 }{2}} (q+ip) ,
\end{equation}
$|n \rangle$ are the Fock states, $\hat{a}^{\dagger} \hat{a}|n \rangle = n |n \rangle$, and $c_n^k = \langle n|\psi_k \rangle$.
The Husimi function provides a picture of the eigenstate in phase space, thus connecting the structure of the state with the different phase-space regions identified in Fig.~\ref{fig02}.

\begin{figure}[h!]
    \centering
\resizebox{\linewidth}{!}{
    \includegraphics{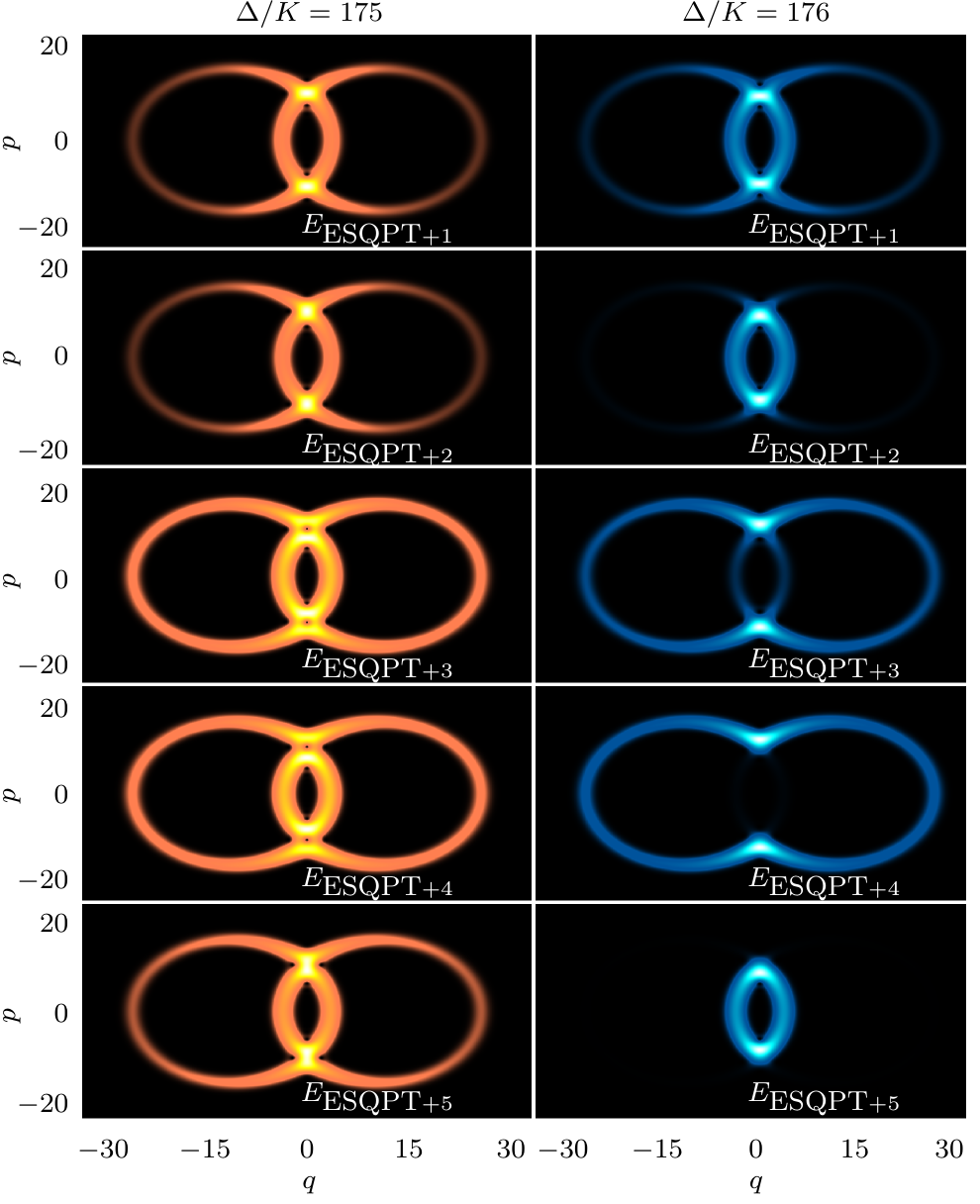}
}
\caption{Husimi function of eigenstates with energy immediately above the ESQPT, for $\Delta/K=175$ (left column) and $\Delta/K=176$ (right column). In all panels: $\epsilon_2/K = 3$.
}
\label{fig04}
\end{figure}

In Fig.~\ref{fig04}, we show the Husimi functions for five eigenstates $\ket{\psi_k}$ with  energy immediately above the energy of the first ESQPT and fixed $\epsilon_2/K=3$. Their energies coincide with those of trajectories in the $\Omega_{\text{in}}$ and $\Omega_{\text{out}}$ regions of the phase space depicted in Fig.~\ref{fig02}. The left column in Fig.~\ref{fig04} is for the odd value $\Delta/K=175$, where avoided crossings occur, and the right column has the even value $\Delta/K=176$, where real crossings occur. For the odd $\Delta/K$, the Husimi function lies in both regions $\Omega_{\text{in}}$ and $\Omega_{\text{out}}$. In contrast, for the even $\Delta/K$, the Husimi function is different from zero either in $\Omega_{\text{in}}$ or in $\Omega_{\text{out}}$, as evident from the two bottom panels on the right side of Fig.~\ref{fig03}.
[When the energy is very close to $E_{\text{ESQPT}}^{\text{(III)}}$, despite $\Delta/K$ being even, the Husimi function in the three top panels on the right is predominantly in one of the two regions, $\Omega_{\text{in}}$ or $\Omega_{\text{out}}$, but has a small probability to be found in the other one.]

In addition to the Husimi functions, we compute the participation ratio of the eigenstates written in the Fock basis,
$\ket{\psi_k} = \sum_{n} c_n^k\ket{n}$. This quantity is defined as
\begin{equation}
\mathcal{PR}^{(k)} = \left(\sum_{n=0}^{N-1} |\langle n|\psi_k \rangle|^4\right)^{-1}.
\end{equation}
It measures the level of delocalization of the state in a chosen basis. The participation ratio is large for an extended state and small for a localized state.

\begin{figure}[h!]
    \centering
    \includegraphics{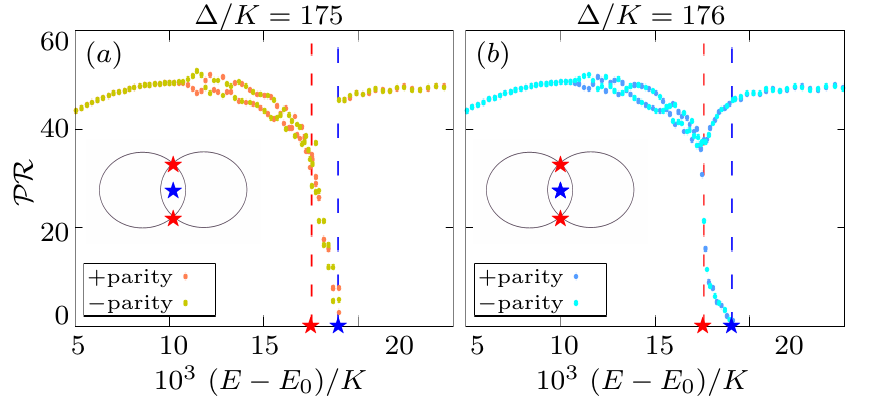}
\caption{Participation ratio in the Fock basis for $\Delta/K=175$ (a) and $\Delta/K=176$ (b). The vertical red dashed line indicates the energy of the first ESQPT (peak in Fig.~\ref{fig01}) and the vertical blue dashed line marks the local maximum energy of the second ESQPT (discontinuous step in Fig.~\ref{fig01}). Colors orange (a) and blue (b) correspond to the positive parity, and colors yellow (a) and cyan (b) to the negative parity.
In both panels: $\epsilon_2/K = 3$.
}
\label{fig05}
\end{figure}

In Fig.~\ref{fig05}, we show the participation ratio for eigenstates in a broad interval of energy, which in addition to the region of crossings, includes also energies smaller than $E^{\text{(III)}}_{\text{ESQPT}}$ and larger than $E^{\text{(III)}}_{\text{step}}$. This allows for a parallel between the features of the full spectrum of Case (III) and the level of delocalization of its  eigenstates. For fixed $\epsilon_2/K=3$, an odd value of $\Delta/K$ is used for Fig.~\ref{fig05}~(a) and an even value for Fig.~\ref{fig05}~(b). In each panel, the points have two different colors, one for the positive parity and the other for the negative parity. In both panels, the vertical red dashed line marks the energy of the first ESQPT, $E_{\text{ESQPT}}^{\text{(III)}}$, and the vertical blue dashed line represents the energy of the second ESQPT at $E_{\text{step}}^{\text{(III)}}$ (see Fig.~\ref{fig01}).

Below the first ESQPT, the behavior of the participation ratio in both panels of Fig.~\ref{fig05} is equivalent, it first grows as the energy increases and then starts decaying at the same time that a peculiar interlace pattern in the participation ratio appears as one approaches $E_{\text{ESQPT}}^{\text{(III)}}$ \cite{footLocESQPT}. The results in both panels are also analogous for energies above $E_{\text{step}}^{\text{(III)}}$, in which case, the participation ratio simply grows as the energy increases.

The distinction between Fig.~\ref{fig05}~(a) and Fig.~\ref{fig05}~(b) happens in the region of the spectrum where the crossings emerge, for $E^{\text{(III)}}_{\text{ESQPT}}<E<E^{\text{(III)}}_{\text{step}}$. 
When $\Delta/K$ is odd (avoided crossings), one sees in Fig.~\ref{fig05}~(a) that the participation ratio decays monotonically as one approaches the local maximum energy at $E^{\text{(III)}}_{\text{step}}$. Recall that classically, this is the energy of the local maximum at $r_0$. In contrast, for even $\Delta/K$, for which the crossings are real, the participation ratio in Fig.~\ref{fig05}~(b) splits in two branches. In one ramification, $\mathcal{PR}$ decreases monotonically, corresponding to the eigenstates whose Husimi functions are distributed over classical trajectories in $\Omega_{\text{in}}$. In the other branch, $\mathcal{PR}$ grows with $E$; these are the eigenstates whose Husimi functions lie in $\Omega_{\text{out}}$.


\section{Quantum tunneling}

\begin{figure*}[t]
    \centering
    \includegraphics{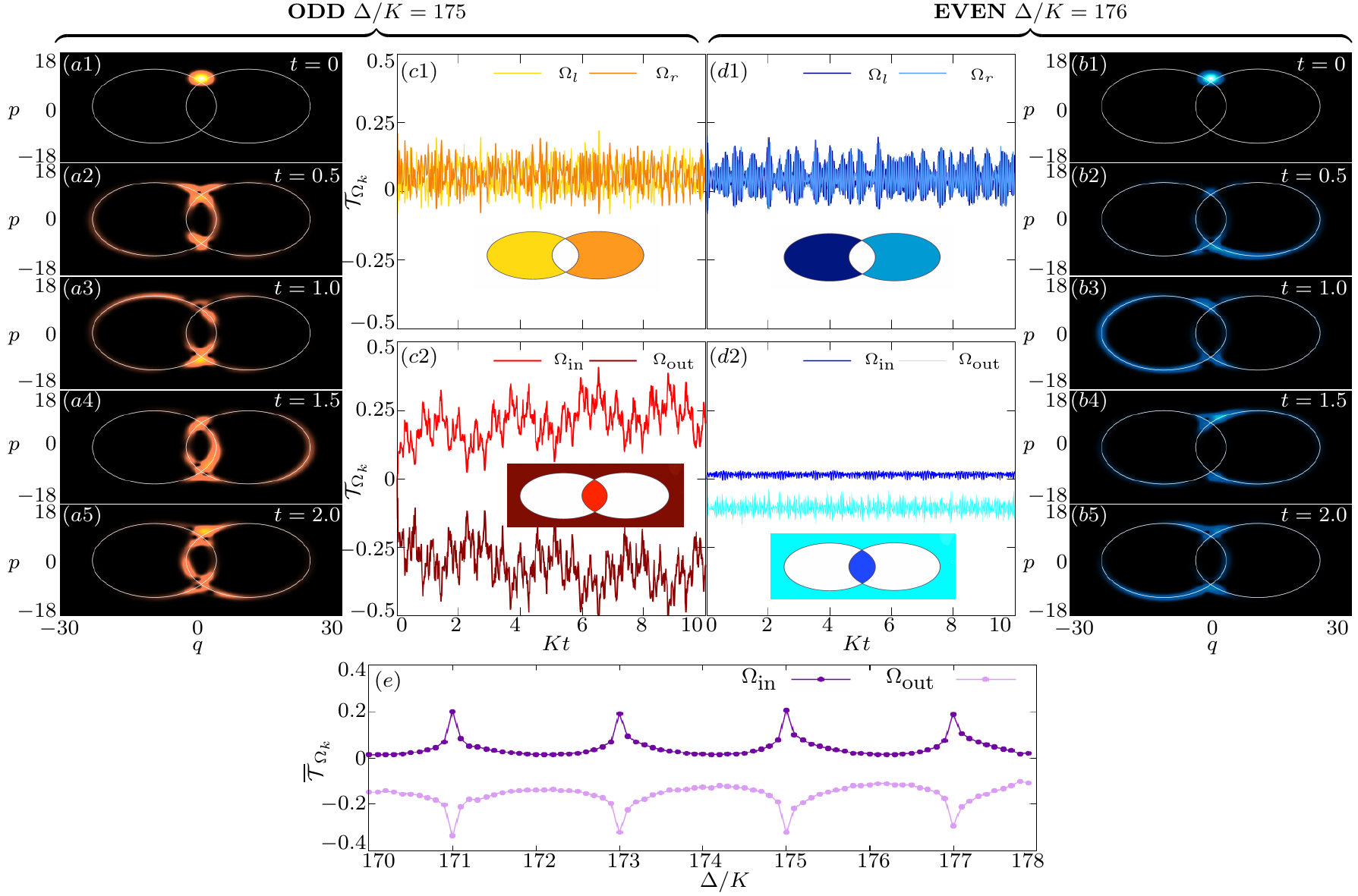}
    \caption{Evolution of the Husimi function, $\mathcal{Q}_{\Psi(t)}^{(q,p)}$, for a coherent state initially centered in the region $\Omega_{\rm{out}}$ for $\Delta/K=175$ (a1)-(a5) and $\Delta/K=176$ (b1)-(b5).
     The white solid line in these panels indicates the separatrix  associated with the hyperbolic fixed points at the energy of the first ESQPT. The energy of the initial coherent state is immediately above the first-ESQPT energy. The results for the effective tunneling $\mathcal{T}_{\Omega_k}$ as a function of dimensionless time are shown for the regions $\Omega_{\rm{l}}$ and for $\Omega_{\rm{r}}$ [(c1) and (d1)] and $\Omega_{\rm{in}}$ and $\Omega_{\rm{out}}$ [(c2) and (d2)], for $\Delta/K=175$ (c1)-(c2) and $\Delta/K=176$ (d1)-(d2). The mean effective tunneling $\overline{\mathcal{T}_{\Omega_k}(t,0)}$ as a function of the parameter $\Delta/K$ is obtained for $Kt\in[0,10]$ in (e). In all panels: $\epsilon_2/K=3$. 
}
    \label{fig06}
\end{figure*}

In this section, we explore the dynamical consequences of the real and avoided crossings of Case (III). Our focus is in the tunneling phenomenon, which involves the quantum mechanical motion of an initial state between regions of phase space that are classically forbidden.
Our analysis requires a quantity capable of measuring how the evolved state samples specific regions of phase space. This can be done with the Husimi function in the position and momentum representation now computed for the evolved state $|\Psi(t)\rangle= e^{-i \hat{H}t} |\Psi (0)\rangle$,
\begin{equation}
\mathcal{Q}_{\Psi(t)}^{(q,p)} = \frac{1}{\pi}\left| \bra{\alpha(q,p)}\ket{\Psi(t)}\right|^2,
\end{equation}
instead of for eigenstates, as in Eq.~(\ref{Eq:HusimiStatic}).

In the following, we describe our method for capturing quantum tunneling to specific regions of phase space. Given a phase space $\mathcal{M}$, let us first define the invariant regions $\Omega$  in which we are interested in studying tunneling dynamics.  We say that a region $\Omega\subset \mathcal{M}$ is invariant if for any initial condition $x\in\Omega$, the classical trajectory stays in that same region $\Omega$, that is, $\varphi_x(t)\in \Omega$ for all $t$, where $\varphi:\mathbb{R}\times \mathcal{M}\rightarrow \mathcal{M}$ is the flow defined by the Hamilton's equations of motion in Eq.~(\ref{Hamil_eq}). We are interested in the four invariant regions of phase space that are depicted in Fig.~\ref{fig02} corresponding to Case (III). Two regions with energy below the energy of the first ESQPT, $\Omega_{\rm{l}}$ and $\Omega_{\rm{r}}$, and two regions with energy above the first ESQPT energy, $\Omega_{\rm{in}}$ and $\Omega_{\rm{out}}$. 

For a particular invariant region in phase space, $\Omega_k\subset \mathcal{M}$, we define the \textit{Husimi volume} at time $t$ as follows~\cite{Villasenor2021},
\begin{equation}
 \label{VQ}
\mathcal{V}_{\Omega_k}\left(t\right)=\iint_{\Omega_k} dqdp \ 
\mathcal{Q}_{\Psi(t)}^{(q,p)},
\end{equation}
which can be evaluated efficiently by random sampling the Husimi function, that is, by Monte Carlo integration \cite{Nader2021}. The Husimi volume is normalized, so that $\mathcal{V}=1$ when the integral in Eq.~(\ref{VQ}) is done over the whole phase space, $\Omega_k=\Omega$.

We also define the \textit{effective tunneling} into/from an invariant region $\Omega_k\subset\mathcal{M}$  from time $t_0$ to time $t$, as
\begin{equation}
\mathcal{T}_{\Omega_k} \! \! \left(t,t_0\right)=\mathcal{V}_{\Omega_k} \! \! \left(t\right)-\mathcal{V}_{\Omega_k} \! \! \left(t_0\right).
\label{Eq:T}
\end{equation}
Notice that if $\mathcal{M}=\bigcup_k\Omega_k$ with $\Omega_k\cap\Omega_{k'}=\emptyset$ for all $k\neq k'$,  as shown in Fig.~\ref{fig02}, then $\sum_k \mathcal{T}_{\Omega_k}\left(t,t_0\right)=0$ for any time $t$. If $\mathcal{T}_{\Omega_k}(t,t_0) > 0$, there exists an effective tunneling towards the region $\Omega_k$, while if $\mathcal{T}_{\Omega_k}(t,t_0) < 0$, it indicates the occurrence of an effective tunneling away from this region. 

The third definition needed for our analysis is the \textit{mean effective tunneling} over the invariant region $\Omega_k$ from time $t_0$ to time $t$, which is written as
\begin{equation}
\overline{\mathcal{T}_{\Omega_k}(t,t_0)}=\frac{1}{t-t_0}
\int_{t_0}^t d\tau \  
\mathcal{T}_{\Omega_k}\left(t,t_0\right).
\label{Eq:AveT}
\end{equation}

In the following two subsections, we present our results for the analysis of quantum tunneling when  $\Delta/K$ takes even or odd values. For the initial states considered in Sec.~\ref{Sec:ODD}, we are able to dynamically determine the values of $\Delta/K$ associated with avoided crossings, and for the initial states in Sec.~\ref{Sec:EVEN} we pinpoint the values of $\Delta/K$ associated with real crossings.

\subsection{Dynamical identification of the parameters for avoided crossings}
\label{Sec:ODD}

We start the study with initial states that have energy between the two ESQPTs, in the energy region of crossings, $E^{\text{(III)}}_{\text{ESQPT}}<E_\Psi<E^{\text{(III)}}_{\text{step}}$.


\subsubsection{Initial coherent state}

In Fig.~{\ref{fig06}}, we consider as initial state, $|\Psi(0)\rangle$, coherent states that are initially centered in the region $\Omega_{\rm{out}}$, as seen in Fig.~{\ref{fig06}}~(a1) and Fig.~{\ref{fig06}}~(b1). The white line in these panels marks the classical separatrix.
The evolution of the Husimi function is shown in Figs.~{\ref{fig06}}~(a1)-(a5) for $\Delta/K=175$ and in Figs.~{\ref{fig06}}~(b1)-(b5) for $\Delta/K=176$. For the odd case, the state spreads in the region $\Omega_{\text{out}}$ and visibly tunnels into the classically forbidden region $\Omega_{\text{in}}$. This contrasts with the even case, where $|\Psi(t)\rangle$ remains predominantly in $\Omega_{\text{out}}$.

The distinctive behavior associated with even and odd $\Delta/K$ is quantified with the effective tunneling $\mathcal{T}_{\Omega_k}\left(t,t_0\right)$ [Eq.~(\ref{Eq:T})] presented in Figs.~{\ref{fig06}}~(c1)-(c2) for $\Delta/K=175$, and in Figs.~{\ref{fig06}}~(d1)-(d2) for $\Delta/K=176$. 
The tunneling into the regions $\Omega_{\text{l}}$ and $\Omega_{\text{r}}$ is minor and analogous for both even and odd $\Delta/K$, as indicated by the very small and quickly saturating positive values of $\mathcal{T}_{\Omega_\text{l}}\left(t,t_0\right)$ and $\mathcal{T}_{\Omega_\text{r}}\left(t,t_0\right)$ in Fig.~{\ref{fig06}}~(c1) and Fig.~{\ref{fig06}}~(d1). Tunneling into $\Omega_{\text{l,r}}$ is simply due to the width of the initial coherent state that is not entirely confined to $\Omega_{\text{out}}$.

More relevant to the discussion are the results in Fig.~{\ref{fig06}}~(c2) and Fig.~{\ref{fig06}}~(d2) corresponding to the regions $\Omega_{\text{in}}$ and $\Omega_{\text{out}}$. For the odd $\Delta/K$ in Fig.~{\ref{fig06}}~(c2), $\mathcal{T}_{\Omega_\text{out}}\left(t,t_0\right)$ [$\mathcal{T}_{\Omega_\text{in}}\left(t,t_0\right)$] is negative [positive], it decreases [increases] on average, and it does not reach a saturation value for the dimensionless time interval considered. This implies that the state  tunnels from $\Omega_{\text{out}}$ into the classically forbidden region $\Omega_{\text{in}}$. In stark contrast with Fig.~{\ref{fig06}}~(c2), $\mathcal{T}_{\Omega_\text{in}}\left(t,t_0\right)$ for the even $\Delta/K$ in Fig.~{\ref{fig06}}~(d2) is very close to zero, indicating nearly no tunneling into $\Omega_{\text{in}}$.

Figure~{\ref{fig06}}~(e) depicts the mean effective tunneling $\overline{\mathcal{T}_{\Omega_k}(t,0)}$ [Eq.~(\ref{Eq:AveT})] as a function of $\Delta/K$, and sums up the results above. This figure provides a tool for the identification of the values of $\Delta/K$ where tunneling into the classically prohibited region $\Omega_{\text{in}}$ is enhanced due to the avoided crossing observed in the spectrum. As evident in the figure, $\Omega_{\text{in}}$ peaks and $\Omega_{\text{out}}$ shows troughs only at odd values of $\Delta/K$. However, the results in Fig.~{\ref{fig06}}~(e) do not single out the even values of $\Delta/K$ as especial cases. To identify these values, one needs to resorts to other initial states, as described later in Sec.~\ref{Sec:EVEN}.


\subsubsection{Quench dynamics}

A scenario that is often explored in nonequilibrium quantum dynamics, which we consider in this subsection, is that of quantum quench dynamics, in which the evolution is started with the abrupt change of a parameter of the Hamiltonian \cite{Mitra2018}.  In our case, the ratio $\Delta/K$ of the Hamiltonian $\hat{H}$ in Eq.~(\ref{Eqeff}) is suddenly modified from $\Delta_0/K$ to $\Delta/K$. The initial state  is prepared in the ground state of $\hat{H}(\Delta_0/K)$ and it evolves according to  $\hat{H}(\Delta/K)$. This provides another example where one can establish through the dynamics that the odd values of $\Delta/K$ are associated with avoided crossings.

\begin{figure}[h]
\hspace{-0.2cm}\includegraphics{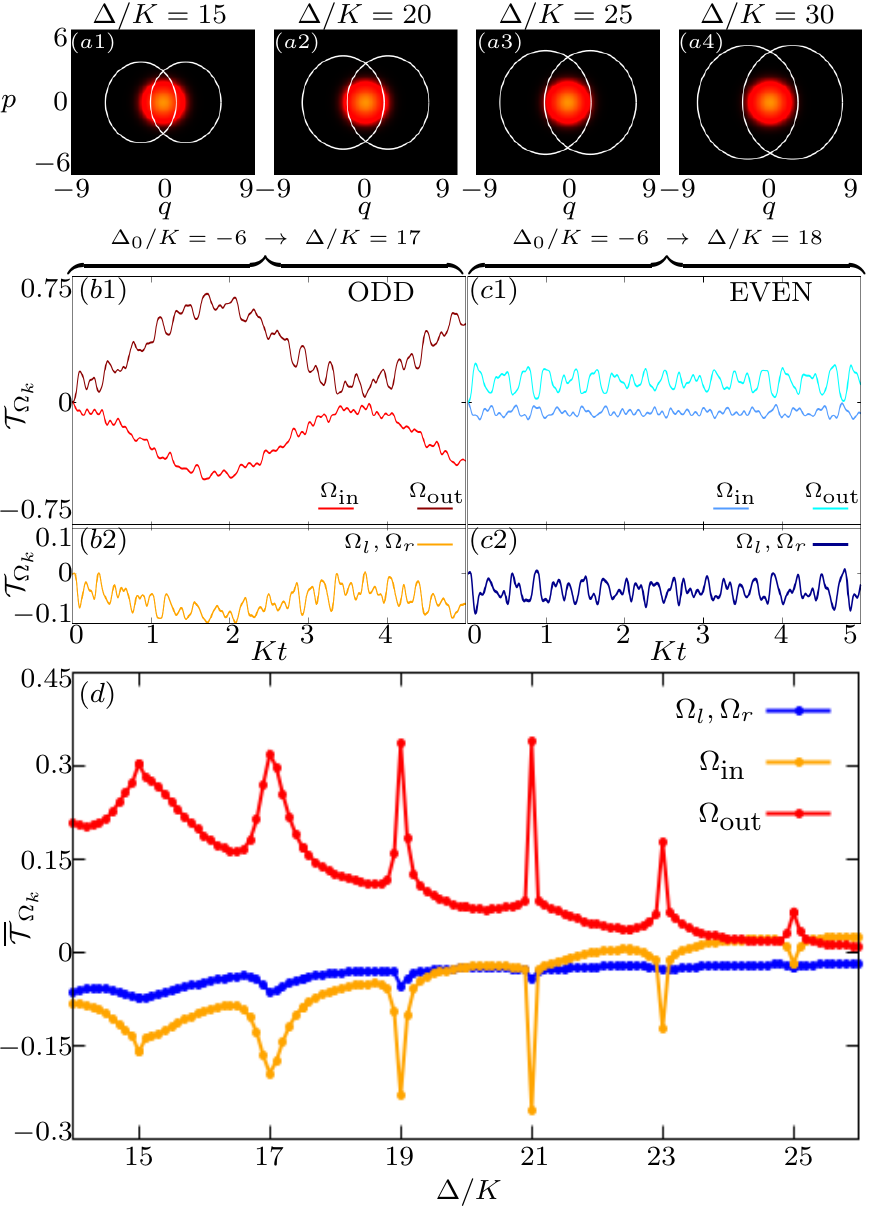}
\vspace*{-0.75cm}
\caption{Husimi function for the ground state of $\hat{H}(\Delta_0/K=-6)$ in the phase space of $H_{cl}(\Delta/K)$ (a1)-(a4). Effective tunneling after  quenching $\Delta_0/K$ to an odd $\Delta/K$ (b1)-(b2) [even $\Delta/K$ (c1)-(c2)]. Mean effective tunneling after the quench as a function of the parameter $\Delta/K$ is obtained for $Kt\in[0,20]$ in (d). In all panels: $\epsilon_2/K=3$.}
\label{fig07}
\end{figure}

We fix $\Delta_0/K$ and investigate the evolution for different values of $\Delta/K$. Our quench is done from Case (I) with $\Delta_0/K=-6$ and $\epsilon_2/K=3$, to Case (III) with $\Delta/K>14$ and $\epsilon_2/K=3$. For our choices, the Husimi function of the initial state is at the center of the phase space, predominantly in the region $\Omega_{\text{in}}$, as seen in Figs.~\ref{fig07}~(a1)-(a4). Notice that as $\Delta/K$ increases, the area within the separatrix of the first ESQPT [white line in Figs.~\ref{fig07}~(a1)-(a4)] grows, causing the initial state to be better confined within $\Omega_{\text{in}}$.

In Figs.~\ref{fig07}~(b1)-(b2) [Figs.~\ref{fig07}~(c1)-(c2)], we show the results of the effective tunneling after the quench to $\Delta/K=17$ [$\Delta/K=18$]. For odd $\Delta/K$, the tunneling from $\Omega_{\rm{out}}$ to $\Omega_{\rm{in}}$ in Fig.~\ref{fig07}~(b1) is significant and more enhanced than in the case of the coherent state in Fig.~\ref{fig06}~(c2). We see an oscillatory behavior associated with the periodic tunneling between the two regions, while $\mathcal{T}_{\Omega_\text{l},\Omega_{r}}$ in Fig.~\ref{fig07}~(b2) remains close to zero.
In contrast, for even $\Delta/K$, the initial state stays mostly confined to $\Omega_{\rm{in}}$, as confirmed by the values of $\mathcal{T}_{\Omega_\text{in},\Omega_\text{out}}$ in Fig.~\ref{fig07}~(c1) and $\mathcal{T}_{\Omega_\text{l},\Omega_{r}}$ in Fig.~\ref{fig07}~(c2), which fluctuate close to zero.

Comparing Fig.~\ref{fig06}~(c2) and Fig.~\ref{fig07}~(b1), one sees that the dynamics is slower for the latter. This is expected, because the energy of the initial state in Fig.~\ref{fig06} is higher than in Fig.~\ref{fig07}.  

Similarly to the analysis in Fig.~\ref{fig06}~(e), the results for the mean effective tunneling in Fig.~\ref{fig07}~(d) corroborates the presence of avoided crossings, and thus enhanced tunneling, for odd values of $\Delta/K$. We notice that the computation of $\overline{\mathcal{T}_{\Omega_k}(t,0)}$ in Fig.~\ref{fig07}~(d) is done using the dimensionless time interval $Kt\in[0,20]$ for all values of $\Delta/K$ presented. However, as $\Delta/K$ increases and the initial state gets deeper confined to $\Omega_{\text{in}}$ [cf. Figs.~\ref{fig07}~(a1)-(a4], tunneling takes longer to happen, which explains why the peaks in Fig.~\ref{fig07}~(d) decrease for larger $\Delta/K$.


\subsection{Dynamical identification of the parameters for real crossings}
\label{Sec:EVEN}

In Fig.~\ref{fig06}~(e), $\overline{\mathcal{T}_{\Omega_\text{in}}(t,0)}$ and $\overline{\mathcal{T}_{\Omega_\text{out}}(t,0)}$ get closest to zero when $\Delta/K$ is even, but there is no particular feature at these points that could indicate anything special about the even values. In this section, we show that one can single out those special even values of $\Delta/K$ by studying the dynamics for initial coherent states centered either in $\Omega_{\text{l}}$ or $\Omega_{\text{r}}$ and placed very close to one of the hyperbolic points. 

\begin{figure*}[t!]
    \centering
    \includegraphics{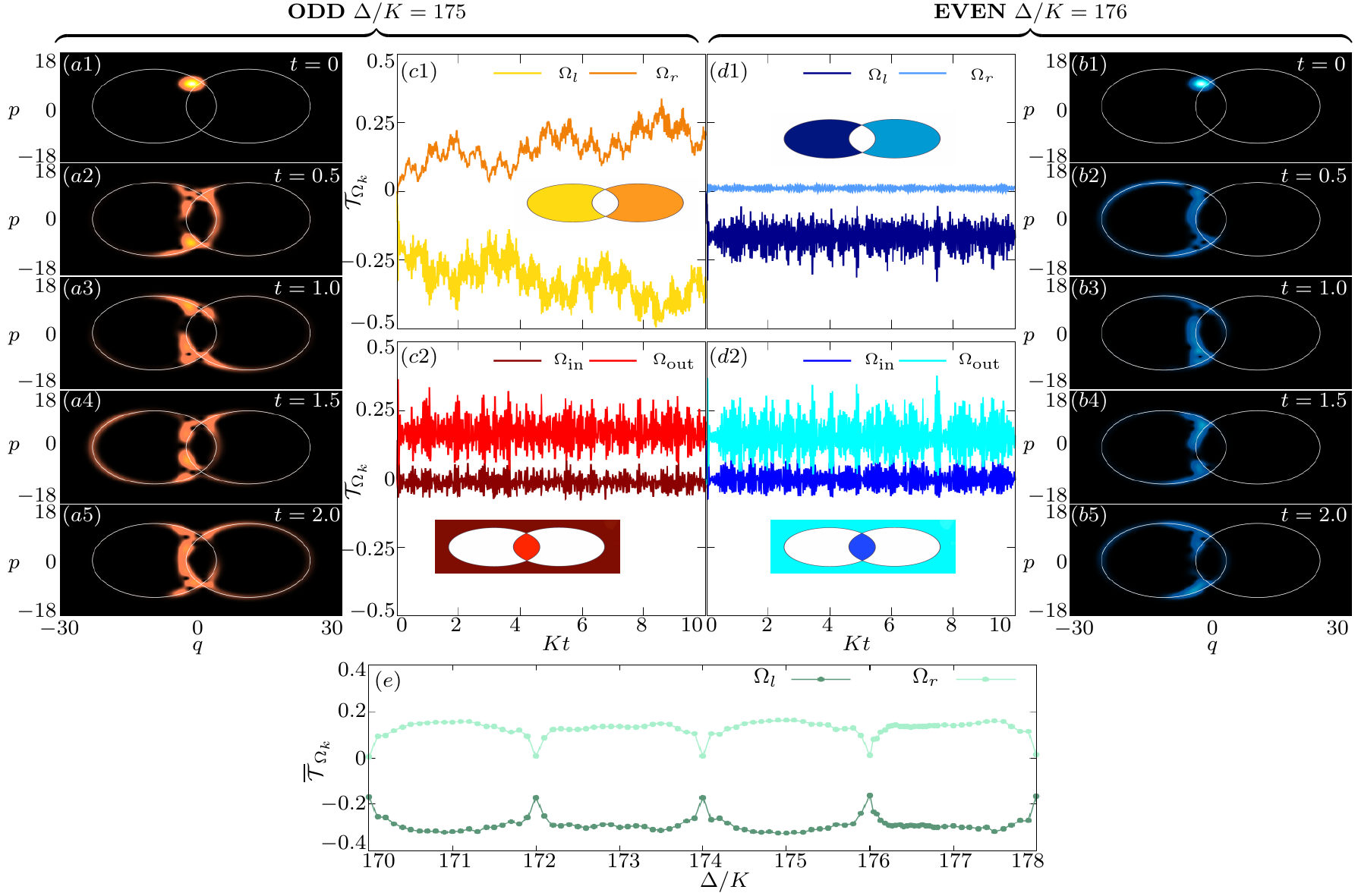}
    \caption{Evolution of the Husimi function, $\mathcal{Q}_{\Psi(t)}^{(q,p)}$, for a coherent state initially centered in the region $\Omega_{\rm{l}}$ for $\Delta/K=175$ (a1)-(a5) and $\Delta/K=176$ (b1)-(b5).
     The white solid line in these panels indicates the separatrix  associated with the hyperbolic fixed points at the energy of the first ESQPT. The energy of the initial coherent state is below the first-ESQPT energy. The results for the effective tunneling $\mathcal{T}_{\Omega_k}$ as a function of dimensionless time are shown for the regions $\Omega_{\rm{l}}$ and for $\Omega_{\rm{r}}$ [(c1) and (d1)] and $\Omega_{\rm{in}}$ and $\Omega_{\rm{out}}$ [(c2) and (d2)], for $\Delta/K=175$ (c1)-(c2) and $\Delta/K=176$ (d1)-(d2). The mean effective tunneling $\overline{\mathcal{T}_{\Omega_k}(t,0)}$ as a function of the parameter $\Delta/K$ is obtained for $Kt\in[0,10]$ in (e). In all panels: $\epsilon_2/K=3$. 
    }
    \label{fig08}
\end{figure*}

Figure~\ref{fig08} is analogous to Fig.~\ref{fig06}, but now for an initial coherent state centered in $\Omega_{\text{l}}$ and close to the hyperbolic point $r_1^{+}$. The Husimi function for odd $\Delta/K$ in Figs.~\ref{fig08}~(a1)-(a5) penetrates into the $\Omega_{\text{r}}$ region, as confirmed by Fig.~\ref{fig08}~(c1), where $\mathcal{T}_{\Omega_\text{l}}$ is negative and $\mathcal{T}_{\Omega_\text{r}}$ is positive. In contrast, for even $\Delta/K$, the region $\Omega_{\text{r}}$ remains mostly empty, as observed in Figs.~\ref{fig08}~(b1)-(b5) and verified with Fig.~\ref{fig08}~(d1). The purpose of Fig.~\ref{fig08}~(c2) and Fig.~\ref{fig08}~(d2) is simply to show that in both cases part of the Husimi function can be found in $\Omega_{\text{out}}$.

The different behaviors for even and odd values of $\Delta/K$ reflect the properties of the spectrum. The initial states in Fig.~\ref{fig08} have energy below but very close to $E^{\text{(III)}}_{\text{ESQPT}}$. They are in the energy region where the degeneracy between the eigenstates in the negative and positive parity sectors gets lifted everywhere, except for the points where $\Delta/K$ is even, as illustrated with arrows in Fig.~\ref{fig03}~(b). This means that one can have tunneling from $\Omega_{\text{l}}$ to $\Omega_{\text{r}}$ for any value of $\Delta/K$, including odd values, as on the left side of Fig.~\ref{fig08}, but excluding even values, as on the right portion of Fig.~\ref{fig08}.

We can use the mean effective tunneling presented in 
Fig.~\ref{fig08}~(e) to determine, directly through the dynamics, the values of $\Delta/K$ where real crossings occur in the spectrum. One see that $\overline{\mathcal{T}_{\Omega_\text{l}}(t,0)}$ exhibits peaks and $\overline{\mathcal{T}_{\Omega_\text{r}}(t,0)}$ exhibits troughs at even values of $\Delta/K$, indicating the suppression of tunneling from the left to the right region at these points. 

Our analysis in Fig.~\ref{fig08}~(e) confirms experimental results obtained in Ref.~\cite{VenkatramanARXIV} for an initial state prepared in one of the wells and with energy close to the ESQPT.  By measuring which-well information after different intervals of time, peaks are found in the relaxation time of this dissipative system for even values of $\Delta/K$, indicating cancellation of tunneling.


\section{Discussion}

 We carried out a detailed analysis of the spectrum and quantum tunneling of the static effective Hamiltonian of a squeeze-driven Kerr oscillator. This is a model with a long history~\cite{Wielinga1993,Cochrane1999,Zhang2017,Puri2017}, that has only recently been realized experimentally in the quantum regime~\cite{Grimm2020,FrattiniPrep,VenkatramanARXIV}. The control parameters considered here and the ones from Ref.~\cite{VenkatramanARXIV}, which are the harmonic coefficient $\Delta$, the nonlinear Kerr coefficient $K$, and the squeezing coefficient $\epsilon_2$. We focused on the parameters values that characterize what we called Case (III), where $\Delta/K > 2\epsilon_2/K$ and the density of states (DOS) exhibits two excited state quantum phase transitions (ESQPTs), one associated with a logarithmic peak and the other with a step discontinuity. 

For even values of $\Delta/K$, real crossings occur for all energy levels, from the ground state energy up to the energy of the second ESQPT at the step discontinuity.
At the intermediate energy regime enclosed by the two ESQPTs, additional avoided crossings emerge for odd values of $\Delta/K$. The condition for the real and avoided crossings in this intermediate energy region was derived from the Einstein-Brillouin-Keller quantization rule. They are related with the existence of pairs of trajectories with the same energy, each trajectory lying in a different invariant region of the classical phase space and covering a different phase-space area.

In the second part of this work, we proposed to use tunneling dynamics to uncover the features of the spectrum that arise in the intermediate energy region between the two ESQPTs. We were able to determine, directly from the dynamics, the special values of $\Delta/K$ related to the energy crossings. This may inspire similar analysis of other driven or non-driven systems described by Hamiltonians comparable to the one considered here.

Our numerical protocol employed the effective flux of the Husimi volume to monitor and quantify the dynamics of an evolved state in specific regions of phase space. We explored two scenarios: quench dynamics and dynamics initiated with a coherent state. Quantum tunneling was controlled by varying $\Delta/K$. It was enhanced for odd values of $\Delta/K$, as we showed for initial states with energy between the ESQPTs, and suppressed for even values of $\Delta/K$, as shown for initial states with energy slightly below the first ESQPT. 

We close this paper with three paragraphs of general discussions.

The first paragraph concerns the generality of our results. Based on our analysis and reviewing the previous literature on one-body systems that exhibit energy crossings, we formulated the following conjecture. For systems with one effective degree of freedom, a sufficient condition for the emergence of energy crossings at an intermediate region of the spectrum and only for specific values of the Hamiltonian parameters, is that the DOS exhibits at least a local logarithmic divergence and a local step discontinuity, which must be consecutive and can 
appear in any order. The crossings are observed at energies between these two transitions. Correspondingly, in the underlying classical system, there should be at least two local stationary points in the phase space, a local hyperbolic point and a local center point, which are related to the logarithmic divergence and the step discontinuity in the DOS, respectively. 

The second paragraph links our findings with current experiments. The results for tunneling involving initial states below (but very close to) the first ESQPT paralleled experimental results obtained in~\cite{VenkatramanARXIV}. In the experiment, a localized state is prepared at the bottom of one of the wells and the activation lifetime for the system to thermalize in between the wells due to dissipation~\cite{Marthaler2006,dykman2012fluctuating} is measured. The incoherent lifetime signal shows peaks that correspond to even values of $\Delta/K$. Our Fig.~\ref{fig08}~(e) shows (inverted) tunneling peaks at the same locations. It thus suggests that the timescale for incoherent flipping between wells is ruled by the tunneling properties of the state close to the ESQPT. From our observations and those made in \cite{FrattiniPrep,VenkatramanARXIV}, we conjecture that the inter-well incoherent dynamics in between wells is determined, only if in part, by the nature of the ESQPT mediating the activation. This conjecture is consistent with the experimental observations made in~\cite{FrattiniPrep} and with the semiclassical and quantum dissipative theory developed therein regarding a new quantum regime of the Arrhenius law, where the incoherent thermalization dynamics between the wells also  mirrors the Hamiltonian spectrum at the level of the ESQPT. Up to now, the theoretical understanding of the incoherent inter-well thermalization dynamics in the driven Kerr oscillator has focused on the engineering of the ground state. Our conjecture suggests the importance of the analysis of the dynamics in the vicinity of ESQPTs. Overall, these findings contribute to the theory of thermal activation in the quantum regime \cite{Hanggi1990,Marthaler2006} and to the engineering of autonomously error-protected bosonic qubits \cite{Puri2017,Grimm2020,FrattiniPrep,VenkatramanARXIV,ruiz2023}.

The third and last paragraph discusses a potential  application of our studies. Experimental systems described by the model Hamiltonian that we considered can serve as a simulation platform of molecular systems, offering control over the strength of interactions that regulate tunneling and non-adiabatic dynamics at avoided crossings and therefore the product yields of reactions. In particular, photochemical reactions are often regulated by real and avoided crossings of electronic excited states, as observed in a wide range of systems, including the photodissociation and geminate recombination of molecules in solution~\cite{Coker1996}, the interfacial electron transfer at functionalized semiconductor interfaces of dye-sensitized solar cells~\cite{sabas2005}, or the photoinduced dynamics of interconversion of chromophores in biological environments~\cite{xin2007}. 


\begin{acknowledgments}
This research was supported by the
NSF CCI grant (Award Number 2124511). FPB wishes to thank funding from the European Union's Horizon
  2020 research and innovation program under grant
  PID2019-104002GB-C21 funded by MCIN/AEI/ 10.13039/501100011033 and,
  as appropriate, by “ERDF A way of making Europe”, by the “European
  Union”, or by the “European Union NextGenerationEU/PRTR”. Computing resources supporting
  this work were partially provided by the CEAFMC and Universidad de Huelva High
  Performance Computer (HPC@UHU) located in the Campus Universitario
  el Carmen and funded by FEDER/MINECO project UNHU-15CE-2848. 
\end{acknowledgments}


\bibliography{biblioNSF2019}

\begin{thebibliography}{58}
\bibitem [{foo()}]{footLocESQPT}%
  \BibitemOpen
  \href@noop {} {}\bibinfo {note} {The localization of the eigenstate at the
  ESQPT was studied for this model with $\Delta=0$ in \cite{ChavezARXIV}. A
  general discussion of this feature for various models was presented before in
  \cite{SantosBernal2015}.}\BibitemShut {Stop}%
\end{thebibliography}

\end{document}